\documentclass{ECOC-author}

\begin{document}

\title{ULTIMATE CAPACITY LIMIT OF A MULTI-SPAN LINK WITH PHASE-INSENSITIVE AMPLIFICATION}

\author{Marcin Jarzyna\ad{1}, Ra\'{u}l Garc\'{\i}a-Patr\'{o}n\ad{2}, Konrad Banaszek\ad{1,3}\corr}

\address{\add{1}{Centre for Quantum Optical Technologies, Uniwersytet Warszawski, Banacha 2c, 02-097 Warszawa, Poland}
\add{2}{Ecole Polytechnique de Bruxelles, CP 165, Universit\'{e} Libre de Bruxelles, 1050 Bruxelles, Belgium }
\add{3}{Wydzia{\l} Fizyki, Uniwersytet Warszawski, Pasteura 5, 02-093 Warszawa, Poland}
\email{k.banaszek@cent.uw.edu.pl}}


\begin{abstract}
The Shannon capacity of a point-to-point link with an optimised configuration of optical amplifiers is compared with general detection strategies permitted by quantum mechanics. Results suggest that the primary application area of receivers based on such strategies may be unamplified short-distance links or free-space optical communication.
\end{abstract}

\maketitle

\section{Introduction}
In the ongoing quest to boost the capacity of optical communication links \cite{Essiambre2012,Bayvel2016}, a relatively unexplored option is to replace conventional detection with more elaborate receivers that operate beyond the standard shot-noise limit. While first designs for such receivers were proposed back in 1970s \cite{Kennedy1973,Dolinar1973}, only the past dozen years have witnessed a number of proof-of-principle experimental demonstrations \cite{Cook2007,Tsujino2011,Chen2012,Muller2012,Becerra2013}. Performance of any type of a receiver is ultimately limited by the Holevo capacity \cite{Holevo1973} which takes into account the most general measurement strategies for the received optical signal that are permitted by quantum mechanics. In contrast, the well established Shannon capacity bound specifically assumes conventional coherent detection of both field quadratures \cite{Essiambre2010}. In the case of a loss-only linear optical channel, the Holevo bound for the spectral efficiency is known to exceed the standard Shannon bound by $1.44~\textrm{bit}/(\textrm{s}\times\textrm{Hz})$ in the bandwidth-limited regime. This figure defines the maximum increase of the information rate that could be attained with unconventional measurement strategies, although designing receivers operating at the Holevo capacity limit remains a dauntingly non-trivial task \cite{Wilde2012}.

The purpose of this work is to assess the possible enhancement of the spectral efficiency beyond the Shannon limit when the propagating signal is regenerated with the help of phase-insensitive optical amplification \cite{Yariv1990}. The locations and the gains of optical amplifiers are numerically optimised under the constraint on the total optical power at any point along the link, using the Holevo capacity bound for the entire channel as the cost function. Such a power constraint can follow e.g.\ from the desire to avoid nonlinear effects in the optical medium. This motivation makes the scenario discussed here distinct compared to the optimisation of the overall energy consumption by an optical transmission system considered recently by Antonelli et al.\ \cite{Antonelli2014}.
It is found that the excess noise introduced by the amplification process closes the gap between the Holevo and the Shannon capacity bounds, leaving rather little room for improvement beyond conventional coherent detection in optically amplified links. This result suggests that the primary area of prospective applications for unconventional receivers may be short-range loss-only links, such as optical interconnects, as well as free-space optical communication, in particular satellite links for which shot-noise limited operation has been experimentally verified \cite{Gunthner2017}.

\section{Capacity limits}

The canonical reference for the capacity limit of a narrowband linear optical channel with additive Gaussian noise follows from the Shannon-Hartley theorem and is given by \cite{Essiambre2010}
\begin{equation}
S_\textrm{Shannon} = \log_2 \left( 1 +\frac{\tau\bar{n}}{1+\nu}\right),
\end{equation}
where $S$ is the spectral efficiency in $\textrm{bits}/(\textrm{s}\times\textrm{Hz})$, $\tau$ is the power loss of the optical channel and $\bar{n}$ is the input signal power spectral density in $\textrm{photons}/(\textrm{s}\times\textrm{Hz})$. The noise term appearing in the denominator is written as a sum of the shot noise of the quadrature measurement, equal to one in the photon number units used here, and the power spectral density $\nu$ in $\textrm{photons}/(\textrm{s}\times\textrm{Hz})$ of the excess noise acquired in the course of propagation.

The above formula relies on the assumption that the readout of field quadratures is performed using conventional shot-noise limited coherent detection. When the most general detection strategies compatible with quantum theory are allowed, the Shannon formula needs to be replaced by the ultimate Holevo capacity limit \cite{Giovannetti2014}
\begin{equation}
S_\textrm{Holevo}=g(\tau\bar{n}+\nu)-g(\nu),
\end{equation}
where $g(x)= \log_2 (1+x) + x \log_2 (1+x^{-1})$. For a loss-only channel with $\nu=0$, the Holevo bound is given by $g(\tau\bar{n})$. In the bandwidth-limited regime when $\tau \bar{n} \gg 1$, the second term in the definition of $g(x)$ can be expanded into a power series in $x=(\tau \bar{n})^{-1} \ll 1$, which yields
 $S_\textrm{Holevo} \approx S_\textrm{Shannon}
+ 1.44~\textrm{bit}/(\textrm{s}\times\textrm{Hz})$.

\begin{figure}
\centering\includegraphics[width=0.8\columnwidth]{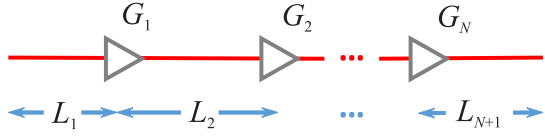}
\caption{A point-to-point communication link including $N$ phase-insensitive amplifiers with gains $G_i$ at relative locations $L_i$.}
\label{Fig:Link}
\end{figure}

\section{Methodology}

We consider a point-to-point optical link extending over a total distance $L$ with uniform attenuation $\alpha$ per unit length. As shown in Fig.~\ref{Fig:Link}, the transmitted signal is regenerated using phase-insensitive amplification at $N$ intermediate nodes labelled with $i=1,\ldots,N$ and located at distances $L_i$ measured with respect to the preceding node. The transformation of the optical field from the channel input to the output of the $i$th amplifier is described by two parameters: $\tau_i$, which specifies the ratio of the signal power right after the $i$th node to the input value, and $\nu_i$, which characterises the exceed noise added by the amplification up to the $i$th node. The recursive relations for the parameters read:
\begin{align}
\tau_i & = G_i \exp(-\alpha L_i) \tau_{i-1}, \nonumber \\
\nu_i & = G_i \exp(-\alpha L_i) \nu_{i-1} + G_i -1. \nonumber
\end{align}
The initial values are $\tau_0=1$ and $\nu_0=0$. Here $G_i$ is the gain of the $i$th amplifier assumed to operate at the quantum limit.
Following the general theory of optical amplification  \cite{Caves1982}, this results in the additive term $G_i-1$ in the equation for $\nu_i$, while the excess noise $\nu_{i-1}$ that has built up until the preceding node is multiplied by the same factor $G_i  \exp(-\alpha L_i)$ as the signal. Taking into account the attenuation over the final unamplified span $L_{N+1}$ between the last regeneration node and the channel output the entire optical channel is characterised by the parameters
\[
\tau = \exp(-\alpha L_{N+1}) \tau_N, \qquad \nu = \exp(-\alpha L_{N+1}) \nu_N.
\]
These values are inserted into the expressions for the Shannon and the Holevo capacity limits. For a given input power spectral density $\bar{n}$ and a number of amplifiers $N$, the power constraint demands that the locations $L_i$ of the nodes and the respective gains $G_i$ need to satisfy the set of inequalities
\[
\tau_i \bar{n} + \nu_i \le \bar{n}, \qquad i=1,\ldots, N
\]
which guarantee that the total optical power does not exceed the input value anywhere along the link.

\begin{figure}
\centering\includegraphics[width=\columnwidth]{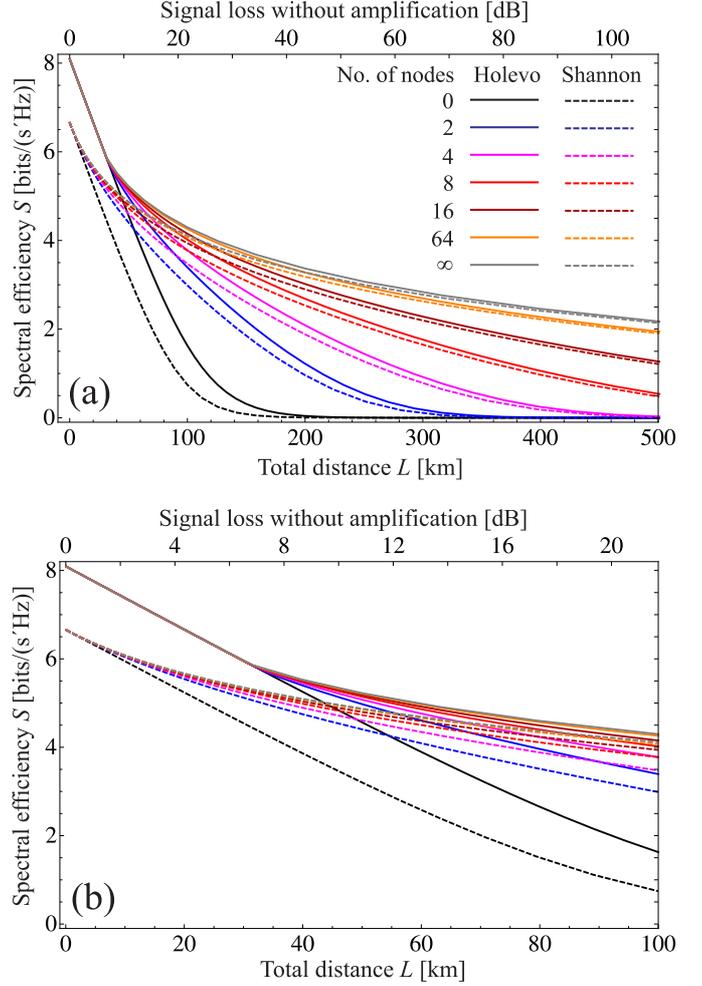}
\caption{(a) The spectral efficiency of a point-to-point link with locations and gains of amplifiers optimised with respect to the Shannon (dashed lines) and the Holevo (solid lines) criterion shown for $N=2,4,8,16,64$ regeneration nodes using colour-coding. The two extreme cases of a loss-only channel and  distributed-amplification cases are depicted with black and grey lines respectively. (b) A close-up for short distances up to $L=100$~km.}
\label{Fig:SE}
\end{figure}

\section{Results}

\subsection{Numerical optimisation}

We carried out numerical optimisation of the locations and the gains of amplifiers for a given number $N$ of regeneration nodes under the constraint that the total power spectral density is less or equal $\bar{n}~\textrm{photons}/(\textrm{s}\times\textrm{Hz})$ at any point of the link. The results of optimisation, carried out independently for the Shannon and the Holevo expressions, are shown in Fig.~\ref{Fig:SE} for the input power spectral density $\bar{n}=100~\textrm{photons}/(\textrm{s}\times\textrm{Hz})$ equivalent to $12.8~\mu\text{W}/\text{THz}$ at the 1550~nm wavelength, the number of nodes $N=2,4,8,16,64$, and the linear attenuation coefficient $\alpha=0.05~\textrm{km}^{-1}$ corresponding to the transmission of the standard SMF-28 fibre at 1550~nm. At $L=0$~km the $1.44~\textrm{bit}/(\textrm{s}\times\textrm{Hz})$ gap in the spectral efficiency is clearly seen.  A qualitative difference between the Shannon and the Holevo criteria is noted for short-haul links: while in the former case it is always beneficial to regenerate the signal using amplification, the Holevo expression is maximised by a loss-only configuration with no need for amplifiers up to 6.73~dB of channel attenuation equivalent to approx.\ $31$~km link span. Beyond that threshold, the gap between the Shannon and the Holevo limits becomes narrower.

\begin{figure}
\centering\includegraphics[width=\columnwidth]{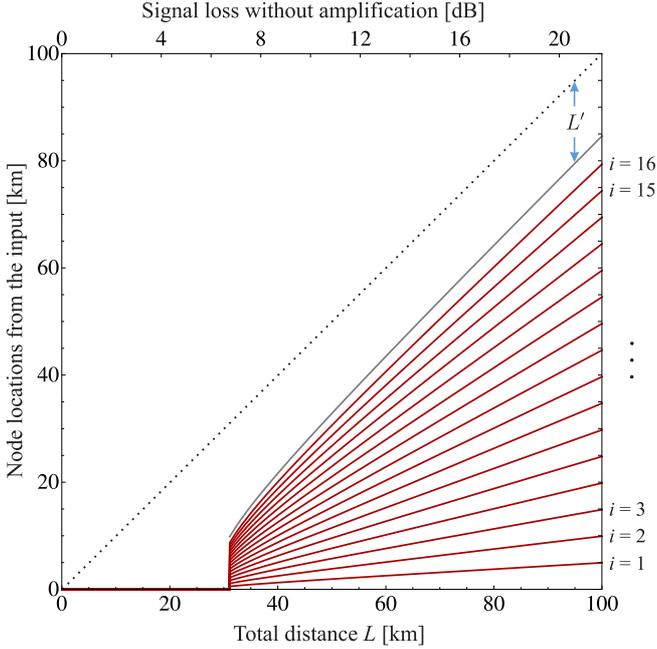}
\caption{Amplifier configuration optimizing the Holevo capacity as a function of the total distance $L$ covered by the link. Colour lines depict locations of individual amplifiers $i=1,2,\ldots,16$ with respect to the input for $N=16$ regeneration stations in total. The solid gray line depicts the optimal termination point of distributed amplification discussed in Sec.~\ref{Sec:DistAmp} with the unamplified last section of the link of length $L'$. The dotted diagonal line serves as a guide to the eye.}
\label{Fig:Locations}
\end{figure}

The general observation following from numerical optimisation is that whenever amplifiers are used, the gain should be set to the maximum value compatible with the total optical power constraint. A distinctive feature has been observed for the optimal locations of the regeneration nodes. When the Shannon criterion is used, all the spans between the nodes are found to be of equal length. As shown in Fig.~\ref{Fig:Locations}, in the case of using the Holevo formula as the optimisation criterion, the last span of the link is to be left unamplified and locations of the regeneration nodes are equidistant over the preceding length of the link. For very long distances the attenuation introduced by the last unamplified span is approx.\ 3~dB.

\subsection{Distributed amplification}
\label{Sec:DistAmp}

The numerical results motivate a model for signal regeneration using distributed optical amplification. In this model, the amplifier gain over an infinitesimal span $dl$ is given by $G(l) \approx 1 + \gamma(l)dl$.
The signal loss $\tau(l)$ and the excess noise $\nu(l)$ become continuous functions of the distance $l$ from the channel input and are governed by a pair of differential equations
\begin{align}
\frac{d\tau}{dl} &= [\gamma(l) - \alpha]\tau(l), \nonumber \\
\frac{d\nu}{dl} & = [\gamma(l) - \alpha]\nu(l) + \gamma(l). \nonumber
\end{align}
If the amplification restores the total optical power to the input level one has $\gamma(l) = \alpha\bar{n}/(1+\bar{n})$ and the analytical solution reads
\[
\tau(l) = \exp[-\alpha l /(1+\bar{n})], \qquad \nu(l) = \bar{n}[1-\tau(l)].
\]
It is seen that compared to the loss-only channel the signal attenuation is reduced by a factor $1/(1+\bar{n})$ and the lost signal power is replaced by the amplification noise. If the Holevo capacity limit is considered, the optimal strategy is to leave unamplified the last section of the link. The length $L'$ of that section can obtained by optimising the Holevo expression with the output signal power spectral density $\exp(-\alpha L')\tau(L-L')\bar{n}$ and the output noise $\exp(-\alpha L')\nu(L-L') $. The optimal termination point $L-L'$ of distributed amplification is shown in Fig.~\ref{Fig:Locations} with a grey line.

\section{Conclusion}
The Holevo capacity limit takes into account the most general detection strategies that go beyond conventional coherent detection of both field quadratures with efficiency  characterised by the Shannon limit. The distinct form of the Holevo expression necessitates a revisited analysis of amplified optical communication links, carried out here under the constraint on the total optical power. It is found that the $1.44~\textrm{bit}/(\textrm{s}\times\textrm{Hz})$ enhancement is material for short distances only, otherwise conventional phase-insensitive amplification combined with coherent detection can deliver performance approaching the Holevo limit. Consequently, unconventional detection strategies can be expected to bring prospectively a meaningful benefit to short-haul pure-loss links such as optical interconnects or in scenarios where signal regeneration is fundamentally impossible, such as satellite optical communication.

\section{Acknowledgements}
We acknowledge insightful discussions with Cristian Antonelli, Saikat Guha, Christoph Marquardt, Antonio Mecozzi, Mark Shtaif, and Jaros{\l}aw P. Turkiewicz, as well as the support of the FNP TEAM project ``Quantum Optical Communication Systems''.

\end{document}